\documentclass[intlimits,a4paper,5p,twocolumn]{elsarticle}

\usepackage{amsmath}
\usepackage{booktabs}
\usepackage{microtype}

\usepackage[utf8]{inputenc}

\journal{Physics Letters B}

\begin{document}

\begin{frontmatter}

  \title{Applicability of pion--nucleus Drell--Yan data in global analysis of nuclear parton distribution functions}

  \author[jy]{Petja Paakkinen\corref{cor1}}
  \ead{petja.paakkinen@jyu.fi}

  \author[jy,hip]{Kari~J. Eskola}
  \ead{kari.eskola@jyu.fi}

  \author[jy,hip,st]{Hannu Paukkunen}
  \ead{hannu.t.paukkunen@jyu.fi}

  \address[jy]{University of Jyvaskyla, Department of Physics, P.O. Box 35, FI-40014 University of Jyvaskyla, Finland}
  \address[hip]{Helsinki Institute of Physics, P.O. Box 64, FI-00014 University of Helsinki, Finland}
  \address[st]{Departamento de F\'{\i}sica de Part\'{\i}culas and IGFAE, Universidade de Santiago de Compostela, E-15782 Galicia, Spain}

  \cortext[cor1]{Corresponding author}

  \begin{abstract}
  Despite the success of modern nuclear parton distribution functions (nPDFs) in describing nuclear hard-process data, they still suffer from large uncertainties. One of the poorly constrained features is the possible asymmetry in nuclear modifications of valence $u$ and $d$ quarks. We study the possibility of using pion--nucleus Drell--Yan dilepton data as a new constraint in the global analysis of nPDFs. We find that the nuclear cross-section ratios from the NA3, NA10 and E615 experiments can be used without imposing significant new theoretical uncertainties and, in particular, that these datasets may have some constraining power on the $u$/$d$ -asymmetry in nuclei.
  \end{abstract}


\end{frontmatter}

\section{Introduction}

Since the discovery of the EMC effect in 1983~\cite{Aubert:1983xm} the nuclear effects in bound-hadron partonic structure have been under active study \cite{Arneodo:1992wf,Malace:2014uea}. For collinearly factorizable hard processes this phenomenon can be described by nuclear modifications of parton distribution functions (PDFs), the latest global extractions being EPS09~\cite{Eskola:2009uj}, DSSZ~\cite{deFlorian:2011fp} and nCTEQ15~\cite{Kovarik:2015cma}, see Refs.~\cite{Eskola:2012rg,Paukkunen:2014nqa} for reviews. Despite the success of nPDFs in describing also nuclear hard-process data from the LHC \cite{Armesto:2015lrg}, they still suffer from large uncertainties. One of the shortcomings is the lack of data which would constrain the nuclear effects of all parton flavours simultaneously without any a priori assumptions. For example, it has been customary to assume that nuclear modifications for both valence quarks $u$ and $d$ are the same. While this assumption has been consistent e.g. with the available LHC data \cite{Armesto:2015lrg} and neutrino-nucleus deep inelastic scattering \cite{Paukkunen:2013grz}, the two are not expected to be exactly the same \cite{Brodsky:2004qa}. It is only recently that an attempt to fit these separately has been carried out \cite{Kovarik:2015cma} but due to the lack of constraining data inconclusive results are obtained. Among other possibilities \cite{Chang:2011ra,Cloet:2012td} it has been also suggested \cite{Dutta:2010pg} that Drell--Yan dilepton data from pion--nucleus collision experiments could be used in nPDF global analyses to constrain the $u$/$d$ -asymmetry. In this Letter, we provide a detailed study of this possibility in terms of the available data and next-to-leading order (NLO) cross-section computations with the EPS09 and nCTEQ15 nPDFs.

\section{Dependence on pion PDFs}

The NA3~\cite{Badier:1981ci}, NA10~\cite{Bordalo:1987cs} and E615~\cite{Heinrich:1989cp} experiments provide pion--nucleus ($\pi^\pm + A$) Drell--Yan dilepton $(l^- l^+)$ production data in the following per-nucleon cross-section ratios:
\begin{align}
  R^{+/-}_A(x_2) &\equiv \frac{\mathrm{d}\sigma(\pi^+ + A \rightarrow l^- l^+ + X) / \mathrm{d}x_2}{\mathrm{d}\sigma(\pi^- + A \rightarrow l^- l^+ + X) / \mathrm{d}x_2}, \\
  R^{-}_{A_1/A_2}(x_2) &\equiv \frac{\frac{1}{A_1} \mathrm{d}\sigma(\pi^- + A_1 \rightarrow l^- l^+ + X) / \mathrm{d}x_2}{\frac{1}{A_2} \mathrm{d}\sigma(\pi^- + A_2 \rightarrow l^- l^+ + X) / \mathrm{d}x_2}.
\end{align}
Here, $x_2 \equiv \frac{M}{\sqrt{s}} \mathrm{e}^{-y}$, where $M$ and $y$ are the invariant mass and rapidity of the lepton pair. The pion--nucleon center-of-mass energy is denoted by $\sqrt{s}$. At leading order (LO), the Drell--Yan cross section reads
\begin{align}
  &\frac{\mathrm{d}\sigma(\pi^\pm + A \rightarrow l^- l^+ + X)}{\mathrm{d}x_2} \\&\overset{\text{LO}}{=} \!\int_{\Delta M}\!\mathrm{d}M \frac{8\pi\alpha^2}{9s x_2 M} \sum_q e_q^2 [q_{\pi^\pm}(x_1) \bar{q}_A(x_2) + \bar{q}_{\pi^\pm}(x_1) q_A(x_2)] \notag ,
\end{align}
where $\alpha$ is the fine-structure constant, $x_1 \equiv \frac{M}{\sqrt{s}} \mathrm{e}^{y} = \frac{M^2}{s x_2}$, and the sum goes over the quark flavors $q$ with $e_q$ being the quark charge.The quark/antiquark distributions in a pion (nucleus) at factorization scale $Q \sim M$ are denoted by $q_{\pi^\pm(A)}$/$\bar{q}_{\pi^\pm(A)}$.

The range of the mass integral $(\Delta M)$ as well as $\sqrt{s}$ depend on the experiment and are $4.1~\mathrm{GeV} < M < 8.5~\mathrm{GeV}$ and $\sqrt{s} = 16.8~\mathrm{GeV}$ for NA3. The NA10 experiment provides data at two different beam energies, 286 GeV ($\sqrt{s} = 23.2~\mathrm{GeV}$) and 140 GeV ($\sqrt{s} = 16.2~\mathrm{GeV}$), with a mass range $4.2~\mathrm{GeV} < M < 15~\mathrm{GeV}$ for the higher and $4.35~\mathrm{GeV} < M < 15~\mathrm{GeV}$ for the lower energy, but in both cases excluding the $\Upsilon$ peak region $8.5~\mathrm{GeV} < M < 11~\mathrm{GeV}$.\footnote{Dutta \textit{et al.}~\cite{Dutta:2010pg} used the NA10 data combined from the two different beam energies. We take these as separate datasets.} In the E615 data the mass range is $4.05~\mathrm{GeV} < M < 8.55~\mathrm{GeV}$ at $\sqrt{s} = 21.7~\mathrm{GeV}$, but with an additional kinematical cut $x_1 > 0.36$, which was imposed by the experiment to reduce contributions from the pion sea quarks.

Assuming the isospin and charge conjugation symmetry we have $u_{\pi^{+}} = d_{\pi^{-}} = \bar{d}_{\pi^{+}} = \bar{u}_{\pi^{-}}$ and $d_{\pi^{+}} = u_{\pi^{-}} = \bar{u}_{\pi^{+}} = \bar{d}_{\pi^{-}}$. Hence, in the limit where the pion sea quarks can be neglected and assuming that the mass integration range is narrow enough so that the scale evolution of the PDFs does not play a role, the LO approximation gives
\begin{align}
  R^{+/-}_A(x_2) &\approx \frac{4\bar{u}_A(x_2) + d_A(x_2)}{4u_A(x_2) + \bar{d}_A(x_2)}, \label{eq:rpmappr} \\
  R^{-}_{A_1/A_2}(x_2) &\approx \frac{4u_{A_1}(x_2) + \bar{d}_{A_1}(x_2)}{4u_{A_2}(x_2) + \bar{d}_{A_2}(x_2)}, \label{eq:rnmappr}
\end{align}
where $u_A$ and $d_A$ are the per-nucleon distributions of $u$ and $d$ quarks in a nucleus $A$ with $Z$ protons,
\begin{align}
  u_A &\equiv \frac{Z}{A} u_{p/A} + \frac{A-Z}{A} d_{p/A},\\
  d_A &\equiv \frac{Z}{A} d_{p/A} + \frac{A-Z}{A} u_{p/A}.
\end{align}
Here, $u_{p/A}$, $d_{p/A}$ are the parton distribution functions of a bound proton and we have again used the isospin symmetry to write $u_{n/A} = d_{p/A}$, $d_{n/A} = u_{p/A}$. As the dependence on the pion PDFs essentially cancels in $R^{-}_{A_1/A_2}$ and $R^{+/-}_A$, these quantities promise to be good candidates for global nPDF analyses, where the objective is to probe the nuclear modifications without being significantly sensitive to (possibly poorly known) pion structure. By comparing Equations~\eqref{eq:rpmappr} and \eqref{eq:rnmappr} we see that while $R^{-}_{A_1/A_2}$ probes dominantly the valence quarks, $R^{+/-}_A$ carries more sensitivity to sea quarks as well.

\begin{figure}[tb!]
  \centering
  \includegraphics[width=\columnwidth]{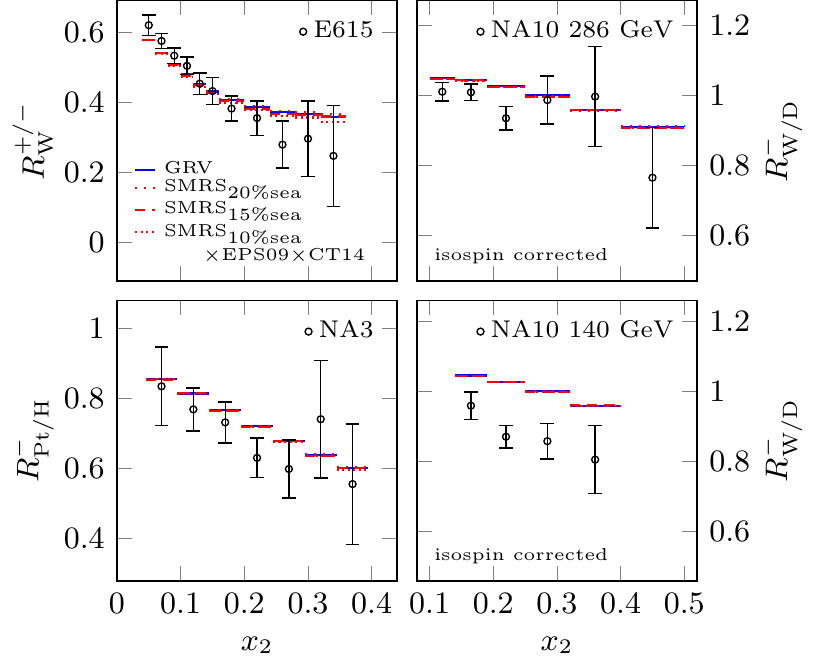}
  \caption{Comparison of NLO predictions with the E615, NA10 and NA3 data. In all panels, we use the GRV (blue) and SMRS (red) PDFs for the pion, and the EPS09 nuclear modifications with the CT14 proton PDFs for the nuclei. In the upper-left panel we have taken into account the kinematical cut $x_1 > 0.36$ and in the right-hand-side panels an isospin correction as described in Section~\ref{sec:isospin} has been applied to the theory predictions.}
  \label{fig:pion_pdfs}
\end{figure}

The above approximative cancellation of the pion PDFs in cross-section ratios has to be tested explicitly in a NLO calculation to avoid including any biased constraints to nPDF analysis. In Figure~\ref{fig:pion_pdfs}, we plot the NA3, NA10 and E615 data along with our NLO results using the GRV~\cite{Gluck:1991ey} and SMRS~\cite{Sutton:1991ay} pion PDFs together with EPS09 nuclear modifications and CT14~\cite{Dulat:2015mca} free-proton PDFs.\footnote{The NA3 data is originally given as $R^-_{\rm H/Pt}$ which we have inverted as it is customary to take the ratio with respect to the lighter nucleus.} For hydrogen and deuterium we use the unmodified CT14 PDFs. In the upper-left panel we have taken into account the kinematical cut $x_1 > 0.36$ and in the right-hand-side panels an isospin correction as described in the next section has been applied. The NLO calculations were done using MCFM~7.0.1~\cite{Campbell:2015qma}. For the data points only statistical errors are available, but these are in any case expected to be dominant in comparison to the systematical errors (except the normalization error of the NA10 data discussed in the next section).

The SMRS pion PDFs provide three different sets to account for the uncertainty in the fraction of pion momentum carried by the sea quarks.We find that the NLO predictions are largely insensitive to the choice of pion PDFs. Especially the SMRS $15\%$ sea set which is to be considered as their central prediction is almost indistinguishable from the GRV results. A slight separation between the different SMRS sets is observed towards large $x_2$ in $R^{+/-}_\text{W}$, but in comparison to the data uncertainties this is insignificant.

\section{Isospin correction and normalization of NA10 datasets}
\label{sec:isospin}

\begin{figure}[tb!]
  \centering
  \includegraphics[width=\columnwidth]{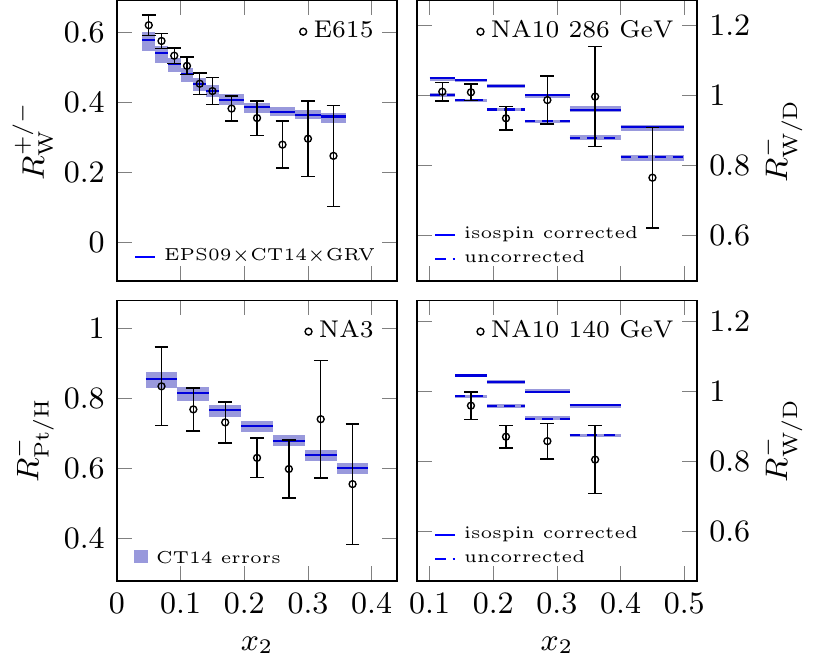}
  \caption{As Figure~\ref{fig:pion_pdfs}, but showing the error estimates from the CT14 PDFs as shaded blue bands for the results obtained with EPS09 and GRV pion PDFs. In the right-hand-side panels we show both the isospin corrected (solid) and uncorrected (dashed) NLO results.}
  \label{fig:proton_pdf}
\end{figure}

The NA10 collaboration has corrected their data for the isospin effects. The exact form of correction was obtained from a LO Monte Carlo simulation but is not quoted point by point along with the data~\cite{Bordalo:1987cs}.\footnote{We thank P.~Bordalo for discussion on this matter.} To mimic these corrections and compare with the data the best we can, we apply an isospin correction by computing the theory predictions as
\begin{equation}
  \begin{split}
    &(R^{-}_\text{W/D})^\text{NLO}_\text{isospin corrected} \\&\qquad\qquad= (R^{-}_\text{isocalar-W/W})^\text{LO}_\text{no nPDFs} \times (R^{-}_\text{W/D})^\text{NLO} ,
  \end{split}
\end{equation}
where ``isoscalar-W" is the isospin-symmetrized W nucleus ($Z=A/2$) and where the LO correction factor $(R^{-}_\text{W/isocalar-W})^\text{LO}_\text{no nPDFs}$ is evaluated with the central set of CT14 without nuclear modifications in PDFs. This correction has been applied on the right-hand-side panels of Figure~\ref{fig:pion_pdfs} and the effect can be seen in Figure~\ref{fig:proton_pdf}, where we plot both the corrected and uncorrected predictions using GRV pion PDFs. In Figure~\ref{fig:proton_pdf}, we also show the error bands from the CT14 proton PDFs (using the asymmetric prescription \cite{Nadolsky:2001yg} to combine the uncertainties from the error sets) which are typically rather small in comparison to the data uncertainties except, perhaps, the E615 data at smallest values  of $x_2$. To some extent, the isospin corrected NA10 data also contain input from the proton PDFs used by the experiment in their Monte Carlo code, but we do not study such a source of uncertainty here further.

We observe that our isospin corrected theory prediction overshoots especially the low-energy NA10 data. This can be accounted for by the systematic overall normalization uncertainty of the data, quoted in~\cite{Bordalo:1987cs} to be $\sigma_{\mathcal{N}^\text{data}} = 6\%$. To compare the predictions from different nPDFs with the NA10 data in shape and not in overall normalization, we normalize the results as follows: We fix the optimal normalization factor $\mathcal{N}^\text{data}$ for each data set and theory prediction separately by minimizing
\begin{equation}
  \chi^2(\mathcal{N}^\text{data}) = \sum_i \frac{(\mathcal{N}^\text{data} R_i^\text{data} - R_i^\text{theory})^2}{(\sigma_i^\text{data})^2} + \frac{(\mathcal{N}^\text{data} - 1)^2}{(\sigma_{\mathcal{N}^\text{data}})^2}
\end{equation}
with respect to data normalization $\mathcal{N}^\text{data}$ \cite{Stump:2001gu}. In the above equation $R_i^{\rm data}$ and $R_i^{\rm theory}$ are the experimental and theoretical values for $i$th bin in a data set, and $\sigma_i^{\rm data}$ is the data uncertainty (here statistical). We then obtain the theory predictions normalized to data as
\begin{equation}
  (R_i^\text{theory})_\text{normalized} = \frac{R_i^\text{theory}}{\mathcal{N}^\text{data}}.
\end{equation}

\begin{table}[tb!]
  \centering
  \caption{Normalization factors for the NA10 data sets.}
  \label{tbl:norm}
  \medskip
  \begin{tabular}{lcc}
  \toprule
  &
  \multicolumn{2}{c}{$\mathcal{N}^\text{data}$}
  \\ \cmidrule(r){2-3}
  nPDF
  & 286 GeV data
  & 140 GeV data
  \\
  \midrule
  EPS09
  & 1.044
  & 1.125
  \\
  nCTEQ15
  & 1.058
  & 1.141
  \\ \bottomrule
  \end{tabular}
\end{table}

\begin{figure}[tb!]
 \centering
 \includegraphics[width=\columnwidth]{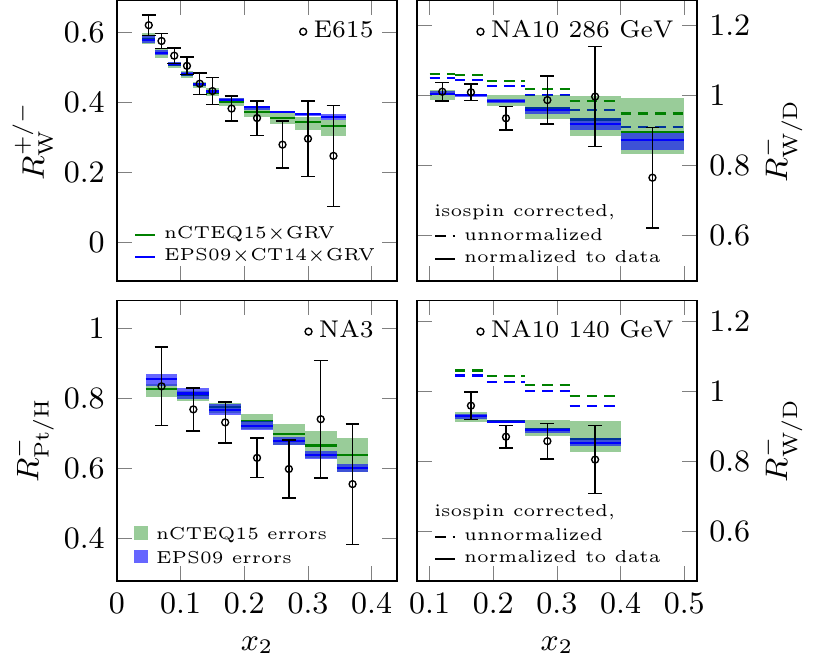}
 \caption{A comparison of the uncertainty bands obtained using the EPS09 (blue lines and bands) and nCTEQ15 (green lines and bands) nuclear PDFs. In the right-hand-side panels we show both the unnormalized (dashed) and results normalized to the data (solid).}
 \label{fig:nuclear_pdfs}
\end{figure}

\noindent The values for $\mathcal{N}^\text{data}$ are given in Table~\ref{tbl:norm} and the normalized results as well as the unnormalized ones are presented in Figure~\ref{fig:nuclear_pdfs} for the EPS09 and nCTEQ15 nuclear PDFs.\footnote{Since nCTEQ15 grids for platinum have not been available for us, we have used their grids for gold instead in $R^{-}_\text{Pt/H}$. Since the mass numbers are very close, $A_\text{Pt} = 195$ and $A_\text{Au} = 197$, this should be an excellent approximation.} For predictions with nCTEQ15 PDFs we use their own free proton set for hydrogen and deuterium (and CT14 for EPS09). When calculating the nPDF errors, we have also normalized each error set separately. We observe that the optimal normalization for the NA10 286 GeV dataset is within the given $6\%$ overall normalization uncertainty, but for the 140 GeV dataset it is more than twice the suggested uncertainty limit. Such a large normalization issue is not unheard of: For example, while the carbon-to-deuteron and lead-to-deuteron nuclear ratios in deep inelastic scattering measured by the E665 collaboration \cite{Adams:1995is} are individually largely apart from other measurements, the lead-to-carbon ratio formed from these two agrees well with other experiments \cite{Arneodo:1996rv}. A similar normalization issue may be in question here as well.

\section{Compatibility with nuclear PDFs}

Comparing the results obtained with the EPS09 and nCTEQ15 nuclear PDFs in Figure~\ref{fig:nuclear_pdfs} we find that both these sets are in a fairly good agreement with the data, but display a large difference in their uncertainty estimates. To understand this, let us study the $R^{-}_\text{W/D}$ ratio measured by NA10. For large $x_2$, only the valence quarks in nuclei contribute and in the LO approximation we have
\begin{equation}
  R^{-}_\text{W/D} \overset{x_2 \rightarrow 1}{\approx} R^\text{W}_\text{V-isoscalar} + R^\text{W}_\text{V-nonisoscalar}, \label{eq:totalR}
\end{equation}
where
\begin{equation}
  R^A_\text{V-isoscalar} \equiv \frac{u^\text{V}_{p/A} + d^\text{V}_{p/A}}{u^\text{V}_{p} + d^\text{V}_{p}}
\end{equation}
is the nuclear modification factor for an average valence quark in an isoscalar nucleus and
\begin{equation}
  R^A_\text{V-nonisoscalar} \equiv \left(\frac{2Z}{A} - 1\right)\frac{u^\text{V}_{p/A} - d^\text{V}_{p/A}}{u^\text{V}_{p} + d^\text{V}_{p}}
\end{equation}
the corresponding non-isoscalarity correction. For neutron-rich nuclei this correction is negative and typically small in comparison to the isoscalar contribution.

\begin{figure}[tb!]
 \centering
 \includegraphics[width=\columnwidth]{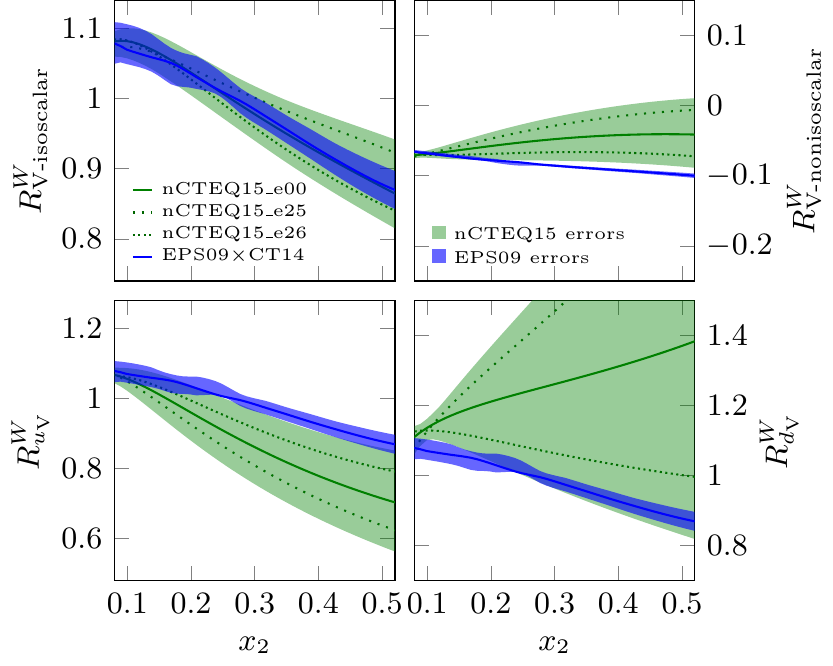}
 \caption{The different LO valence-quark contributions to $R^{-}_\text{W/D}$ (upper panels) and the valence quark nuclear modification factors (lower panels) at factorization scale $Q = 5~\mathrm{GeV}$. Solid lines correspond to the EPS09 (blue) and nCTEQ15 (green) central sets and dotted lines indicate the error sets 25 and 26 of the nCTEQ15. The uncertainty bands are shown as light green (nCTEQ15) and light blue (EPS09) bands.}
 \label{fig:nuclear_valence_mods}
\end{figure}

In Figure~\ref{fig:nuclear_valence_mods}, we plot these two components for tungsten along with the nuclear modification factors
\begin{equation}
  R^{W}_{u_\text{V}} \equiv \frac{u^\text{V}_{p/A}}{u^\text{V}_{p}}, \qquad R^{W}_{d_\text{V}} \equiv \frac{d^\text{V}_{p/A}}{d^\text{V}_{p}}
\end{equation}
at factorization scale $Q = 5~\mathrm{GeV}$. We find that EPS09 and nCTEQ15 agree on $R^\text{W}_\text{V-isoscalar}$, which is well constrained in both analyses, but there is a slight disagreement on $R^\text{W}_\text{V-nonisoscalar}$. In addition, we see that nCTEQ15 has significantly larger error bands in both of these components. To study this difference in more detail, we plot in Figure~\ref{fig:nuclear_valence_mods} also the nCTEQ15 error sets 25 and 26, which give the largest deviations from the central-set predictions. We can make two observations: First, from the lower panels in Figure~\ref{fig:nuclear_valence_mods}, we see that these two error sets are related to the nuclear modifications of $u$ and $d$ valence quarks with set 25 giving the most extreme difference, and set 26 being closer to uniform modifications. Second, from the upper panels in Figure~\ref{fig:nuclear_valence_mods}, we find that the deviations from the central prediction are in the same direction for both $R^\text{W}_\text{V-isoscalar}$ and $R^\text{W}_\text{V-nonisoscalar}$ (upwards for set 25, downwards for set 26), and combine additively in Equation~(\ref{eq:totalR}) thereby explaining the larger error bands seen in Figure~\ref{fig:nuclear_pdfs}.

\begin{figure}[tb!]
 \centering
 \includegraphics[width=\columnwidth]{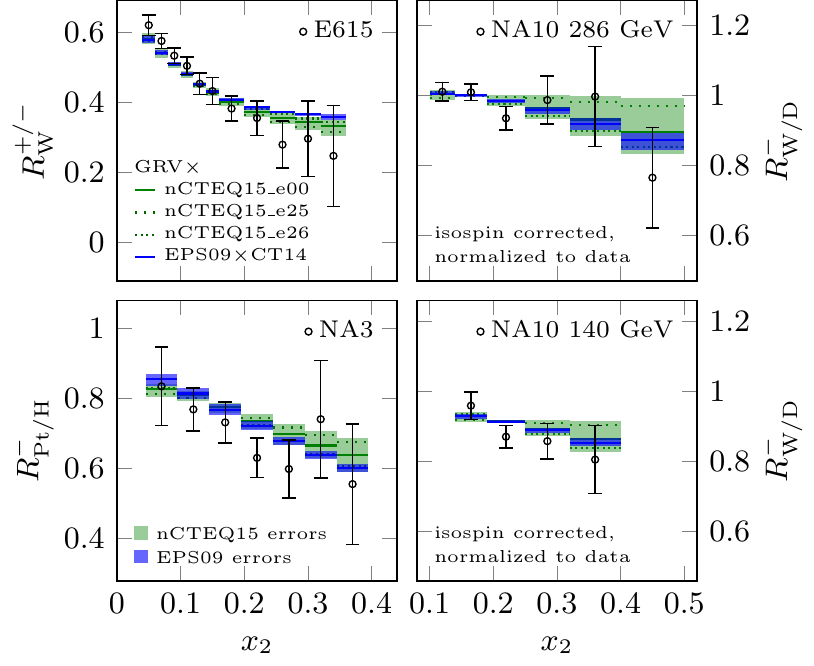}
 \caption{As Figure~\ref{fig:nuclear_pdfs}, but with only normalized results shown and the nCTEQ15 error sets 25 and 26 (dotted lines) plotted.}
 \label{fig:nuclear_pdfs_final}
\end{figure}

It is now evident that the studied observables are sensitive to the mutual differences between $u$ and $d$ valence quark nuclear modifications. On one hand, the EPS09 error sets underestimate the true uncertainty because flavor dependence of valence quark nuclear modifications was not allowed in that particular analysis. On the other hand, the nCTEQ15 error bands are large since the flavor dependence was allowed, but not well constrained in their analysis. The size of nCTEQ15 error bands suggest that the pion--nucleus Drell--Yan data can have some constraining power on the difference of valence modifications. Indeed, in Figure~\ref{fig:nuclear_pdfs_final} we plot the predictions using the nCTEQ15 error sets 25 and 26, and observe that the most extreme deviation from identical nuclear modifications of $u$ and $d$ quarks given by set 25 is disfavored by NA3 and NA10 data.

\begin{figure}[tb!]
 \centering
 \includegraphics{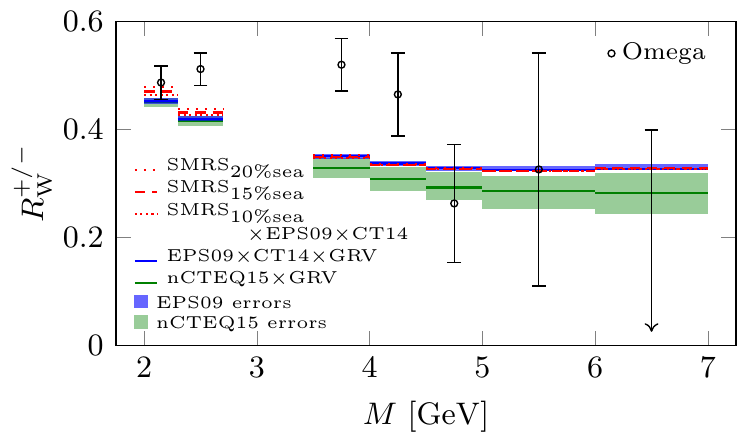}
 \caption{Comparison of the Omega data with predictions using the GRV (blue) and SMRS (red) pion parton distributions together with the EPS09 nuclear modifications combined to the CT14 proton PDFs and also from using the nCTEQ15 (green) nuclear PDFs with the GRV pion PDFs.}
 \label{fig:omega}
\end{figure}

In addition to the NA3, NA10 and E615 data we have studied also the results from the Omega experiment~\cite{Corden:1980xf}. The data at $\sqrt{s} = 8.7~\mathrm{GeV}$ as a function of the lepton pair invariant mass are shown in Figure~\ref{fig:omega} for $x_\mathrm{F} \equiv \frac{2p_\mathrm{L}^*}{\sqrt{s}} > 0$, where $p_\mathrm{L}^*$ is the longitudinal momentum of the lepton pair along the beam line in the center-of-mass frame. We find that the data disagree with theory predictions in bins around the J/$\psi$ peak. Furthermore, at low invariant masses the choice of pion PDFs becomes significant and that especially towards larger invariant masses the data are not precise enough to discriminate between the nuclear PDFs. Hence it is not reasonable to include this dataset into a global nPDF analysis.

\section{Conclusions}

We have studied the prospects of including NA3, NA10, E615 and Omega pion--nucleus Drell--Yan data to global analyses of nuclear parton distribution functions. The NA3, NA10 and E615 data are compatible (modulo NA10 normalization at lower beam energies) with modern nPDFs and can thus be used in a global analysis without causing significant tension. The Omega data is not compatible with the NLO theory predictions and not precise enough to be useful in the nPDF analysis. The cross-section ratios used in the experiments are largely independent of pion parton distributions and hence the inclusion of these data will not impose significant new theoretical uncertainties to the analysis. Some sensitivity to baseline proton PDFs however still persists. When implementing these data to a global analysis, one needs to take into account the isospin correction and normalization uncertainty in the NA10 datasets. This can be done as described above. These pion--nucleus Drell--Yan data will be included in the successor of the EPS09 analysis~\cite{EPPS16}.

The considered nuclear ratios are sensitive to the possible $u$/$d$ -asymmetry of nuclear modification factors but the data are not precise enough to pin down this difference completely. Regarding this matter we seem to reach a somewhat different conclusion than Dutta \textit{et al.}~\cite{Dutta:2010pg} who claimed that NA3 data would favor flavor-dependent nuclear PDFs. We, in our analysis, find a very good agreement between the data and $u$/$d$ -symmetric (EPS09) nuclear modifications. Moreover, our analysis suggests that the most extreme differences in $u$ and $d$ quark nuclear modifications as given by particular nCTEQ15 error sets are disfavored by the NA3 and NA10 datasets.

\section*{Acknowledgements}

This research was supported by the Academy of Finland, Project 297058 of K.J.E., and by the European Research Council grant HotLHC ERC-2011-StG-279579 and by Xunta de Galicia (Conselleria de Educacion) -- H.P. is part of the Strategic Unit AGRUP2015/11. P.P. gratefully acknowledges the financial support from the Magnus Ehrnrooth Foundation.

\bibliographystyle{elsarticle-num}
\bibliography{piA-DY-article}

\end{document}